\documentclass[journal,letterpaper]{IEEEtran_nssmic}
\usepackage{graphicx}
\usepackage{epsfig}
\begin{document}
%
\title{The ATLAS Pixel Detector}
\author{Fabian~H\"{u}gging
\thanks{Manuscript received November 15, 2004; revised August 5,
2005.}
\thanks{This manuscript is submitted by Fabian~H\"{u}gging on behalf of the ATLAS Pixel Collaboration.}
\thanks{F. H\"{u}gging is with Physikalisches Institut, Universit\"{a}t Bonn,
  Nussallee 12, D-53115 Bonn, Germany (telephone: +49-228-73-3210, e-mail: huegging@physik.uni-bonn.de).}}

\maketitle

\begin{abstract}
The ATLAS Pixel Detector is the innermost layer of the ATLAS
tracking system and will contribute significantly to  the ATLAS
track and vertex reconstruction. The detector consists of
identical modules, arranged in three barrels concentric with the
beam line and centered on the interaction point and three disks on
either side for the forward region.

The position of the Pixel Detector near the interaction point
requires excellent radiation hardness, mechanical and thermal
robustness and good long-term stability, all combined with a low
material budget. The detector layout, results from production
modules and the status of assembly are presented.
\end{abstract}


\section{Introduction}

\PARstart{T}{he} ATLAS Inner Detector~\cite{IDTDR} is designed for
precision tracking of charged particles with 40~MHz bunch crossing
identification. It combines tracking straw tubes in the outer
transition-radiation tracker (TRT), the microstrip detectors of
the semiconductor tracker (SCT) at intermediate radii with the
Pixel Detector, the crucial part for vertex reconstruction, as the
innermost component.

The Pixel Detector~\cite{PDTDR} is subdivided into three barrel
layers and three disks on either side for the forward direction.
The innermost barrel layer is close to the beam pipe at radius
$r=50.5$~mm, the other two layers are at $r=88.5$~mm and
$r=122.5$~mm. With a total length of approx. 1.3~m this layout
results in a three-hit system for particles with $|\eta|<2.5$.

The main components are approx.~1700 identical modules,
corresponding to a total of $8\cdot 10^7$ pixels. The modules
consist of a package composed of sensors and readout-chips mounted
on a hybrid. They have to be radiation hard to an ATLAS life time
dose of 50~MRad or $10^{15}$ neutron-equivalent.

\section{Module Layout}

\begin{figure}[htb]
    \centerline{\includegraphics[width=.9\columnwidth]{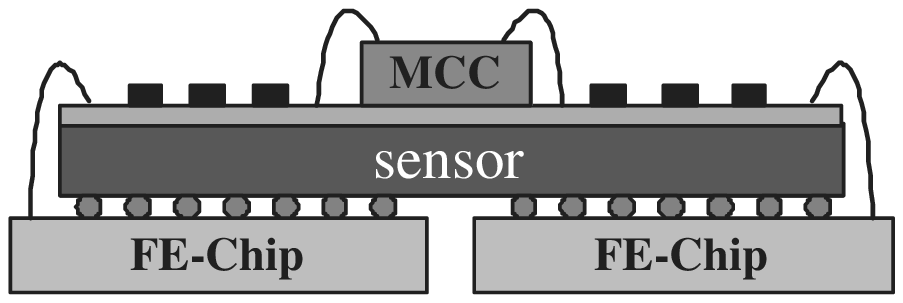}}
 \caption{Cross-section of an ATLAS pixel module.\label{fig:x-sec}}
\end{figure}

A pixel module consists of an oxygenated single n-on-n silicon
sensor, approx.~2$\times$6~cm$^2$ in size~\cite{sensor}. The
sensor is subdivided into 47,268 pixels which are connected
individually to 16 front-end (FE) chips using fine pitch ``bump
bonding'' either done with Pb/Sn by IZM\footnote{Institut f\"ur
Zuverl\"assigkeit und Mikrointegration, Berlin, Germany.} or with
Indium by AMS\footnote{Alenia Marconi Systems, Roma, Italy.}.
These chips are connected to a module-control chip
(MCC)~\cite{MCC} mounted on a kapton-flex-hybrid glued onto the
back-side of the sensor. The MCC communicates with the
off-detector electronics via opto-links, and power is fed into the
chips via cables connected to the flex-hybrid. A cross-section of
an ATLAS pixel module is shown in figure~\ref{fig:x-sec}.

To provide a high space-point resolution of approx.~12$\,\mu$m in
azimuth ($r\phi$), and approx.~110$\,\mu$m parallel to the LHC
beam ($z$), the sensor is subdivided into 41,984 ``standard''
pixels of 50~$\mu$m in $r\phi$ by 400~$\mu$m in $z$, and 5284
``long'' pixels of $50 \times 600$~$\mu$m$^2$. The long pixels are
necessary to cover the gaps between adjacent front-end chips. The
module has 46,080 read-out channels, which is smaller than the
number of pixels because there is a 200~$\mu$m gap in between FE
chips on opposite sides of the module, and to get full coverage
the last eight pixels at the gap must be connected to only four
channels (``ganged'' and ``inter-ganged'' pixels). Thus  on 5\% of
the surface the information has a two-fold ambiguity that will be
resolved off-line.

The FE chips~\cite{fabian} built in the IBM $0.25\,\mu$m
technology contain 2880 individual charge sensitive analogue
circuits with a digital read-out that operates at 40~MHz. The
analogue part consists of a high-gain, fast preamplifier followed
by a DC-coupled second stage and a differential discriminator. The
threshold of the discriminator ranges up to 1~fC, its nominal
value being 0.5~fC. When a hit is detected by the discriminator
the pixel address is provided together with the time over
threshold (ToT) information which allows reconstruction of the
charge seen by the preamplifier.

\section{Module Performance}

During prototyping several prototype modules have been built with
two generations of radiation-hard chips in $0.25\,\mu$m-technology
before the production started with the final chip generation in
early 2004. Up to now roughly 200 modules have been built; in
order to assure full functionality of the modules in the
experiment, each module will be extensively tested after assembly
including measurements at the production sites before and after
thermal cycling. Moreover, several modules from different
production sites have been tested in a test beam and after
irradiation with charged hadrons, the later includes lab tests as
well as test beam studies with irradiated modules.

\subsection{Laboratory measurements}
\label{sec:lab}
\begin{figure}[htb]
\centerline{\includegraphics[width=.8\columnwidth]{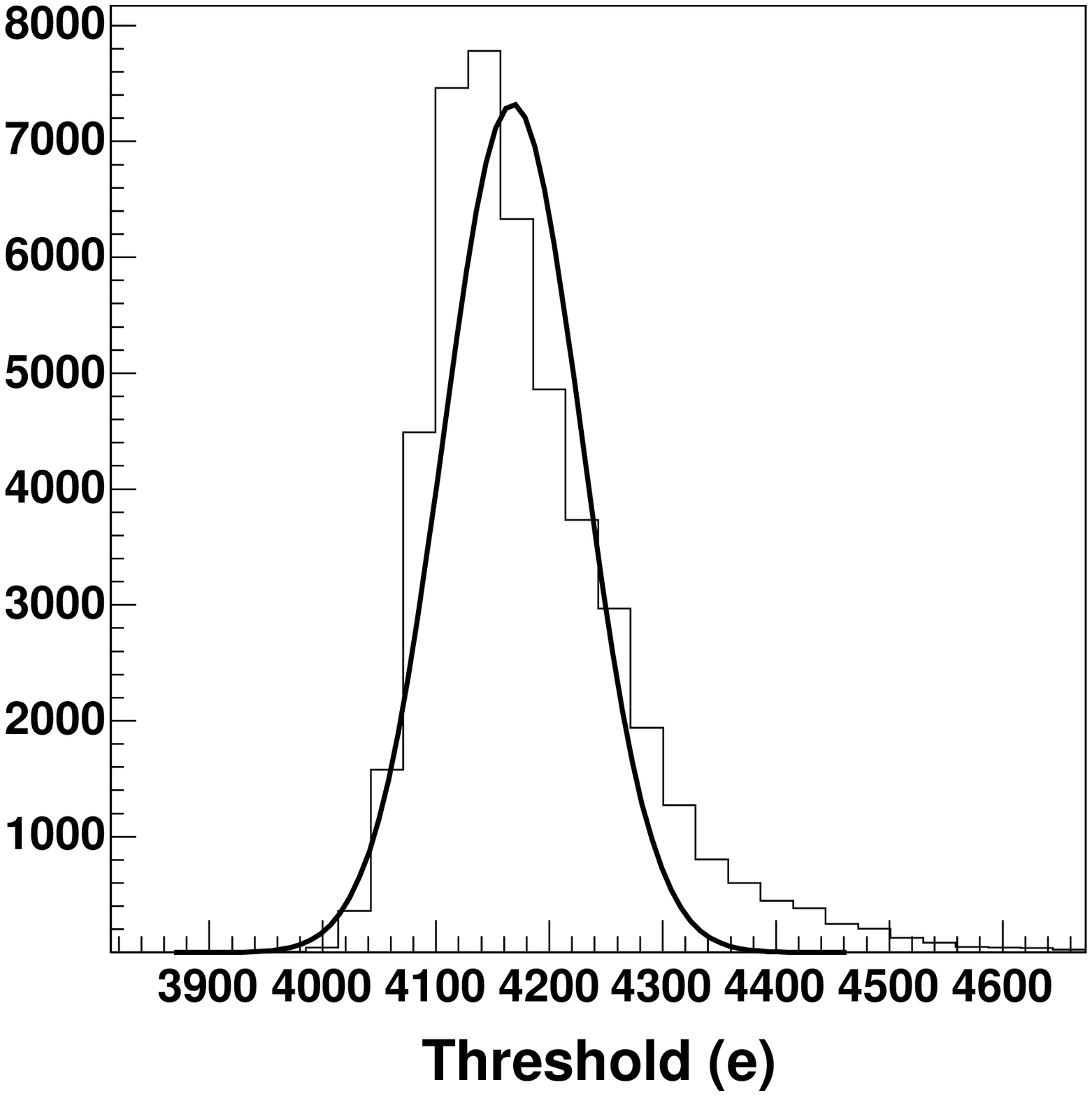}}
\caption{Threshold distribution of a module after the tuning
procedure, the mean threshold is 4,170~e$^{-}$ with a dispersion
of 61~e$^{-}$. \label{fig:threshold}}
\end{figure}

\begin{figure}[t]
\centerline{\includegraphics[width=.9\columnwidth]{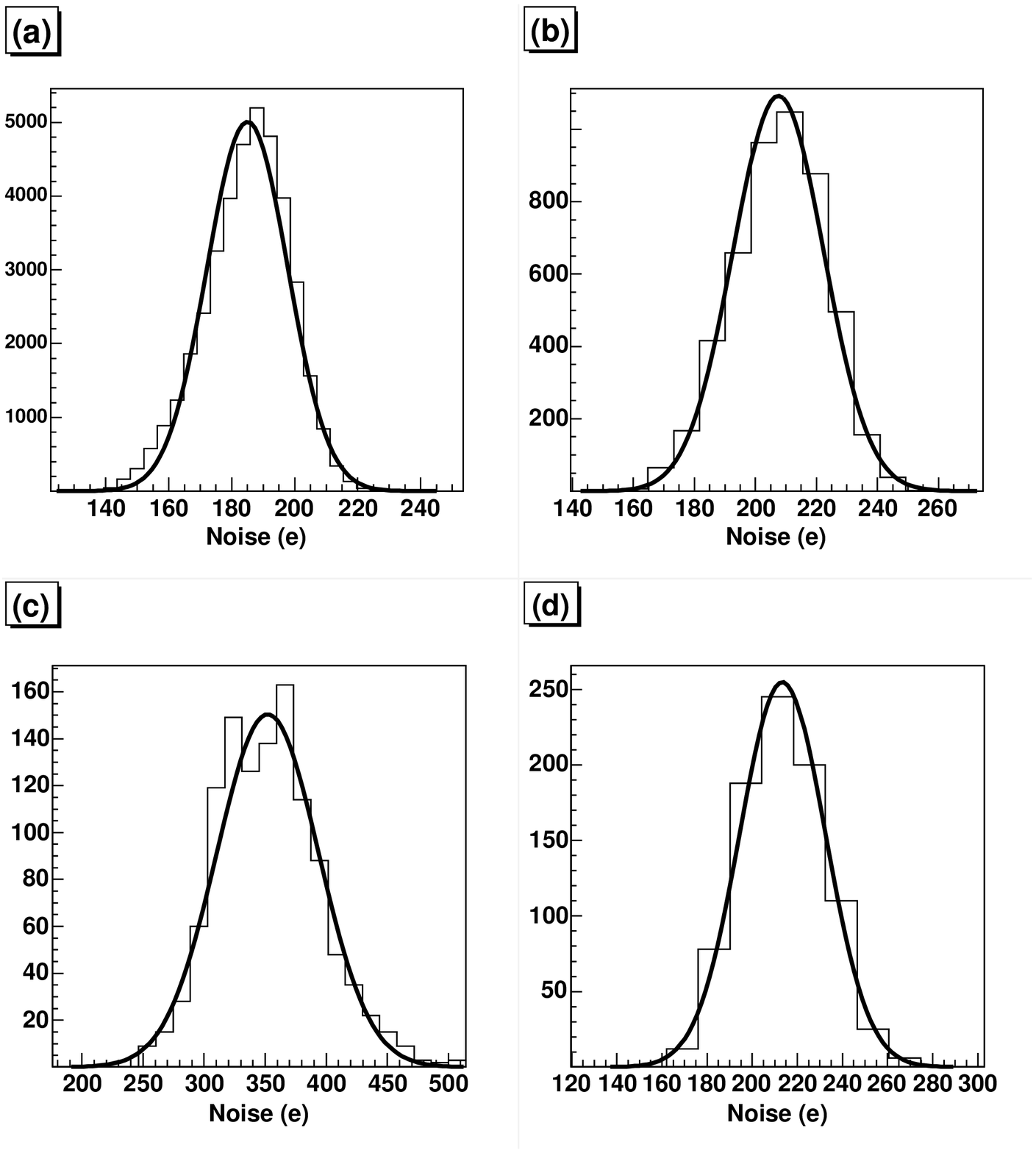}}
\caption{Noise distributions for the different pixel types of a
module after the tuning procedure, (a) for standard pixel with a
mean noise of 185~e$^{-}$ and $\sigma=13~$e$^{-}$, (b) long pixel
with a mean noise of 208~e$^{-}$ and $\sigma=15~$e$^{-}$, (c)
ganged pixel with a mean noise of 352~e$^{-}$ and
$\sigma=42~$e$^{-}$ and (d) inter-ganged pixel with a mean noise
of 213~e$^{-}$ and $\sigma=19~$e$^{-}$. \label{fig:noise}}
\end{figure}

 An important test that allows a large range of
in-laboratory measurements is the threshold scan. Signals are
created with on-chip charge injection for each pixel individually.
Scanning the number of hits versus the so injected charge yields
the physical value of the threshold of the discriminator and the
equivalent noise charge as seen by the preamplifier. A set of such
scans is used to reduce the threshold dispersion by adjusting a
7-bit DAC-parameter individually for each channel, a procedure
that takes about 1 hour. The resulting threshold and noise after
threshold tuning is shown in figures~\ref{fig:threshold}
and~\ref{fig:noise}. Typically approx.~60~e$^{-}$ threshold
dispersion across a module and a noise value of below 200~e$^{-}$
for standard pixels is achieved, as is needed for good
performance.

Note that there is not a single pixel with threshold lower than
$3,900~e^{-}$. This shows the high tuning capability of this chip
allowing to reach small thresholds on the whole module without any
pixel having its threshold too close to the noise, a fact of
particular importance after irradiation. We also measured the
cross-talk to a few per cent for standard $50\times
400$~$\mu$m$^2$ pixels.

\begin{figure}[b]
\centerline{\includegraphics[width=.9\columnwidth]{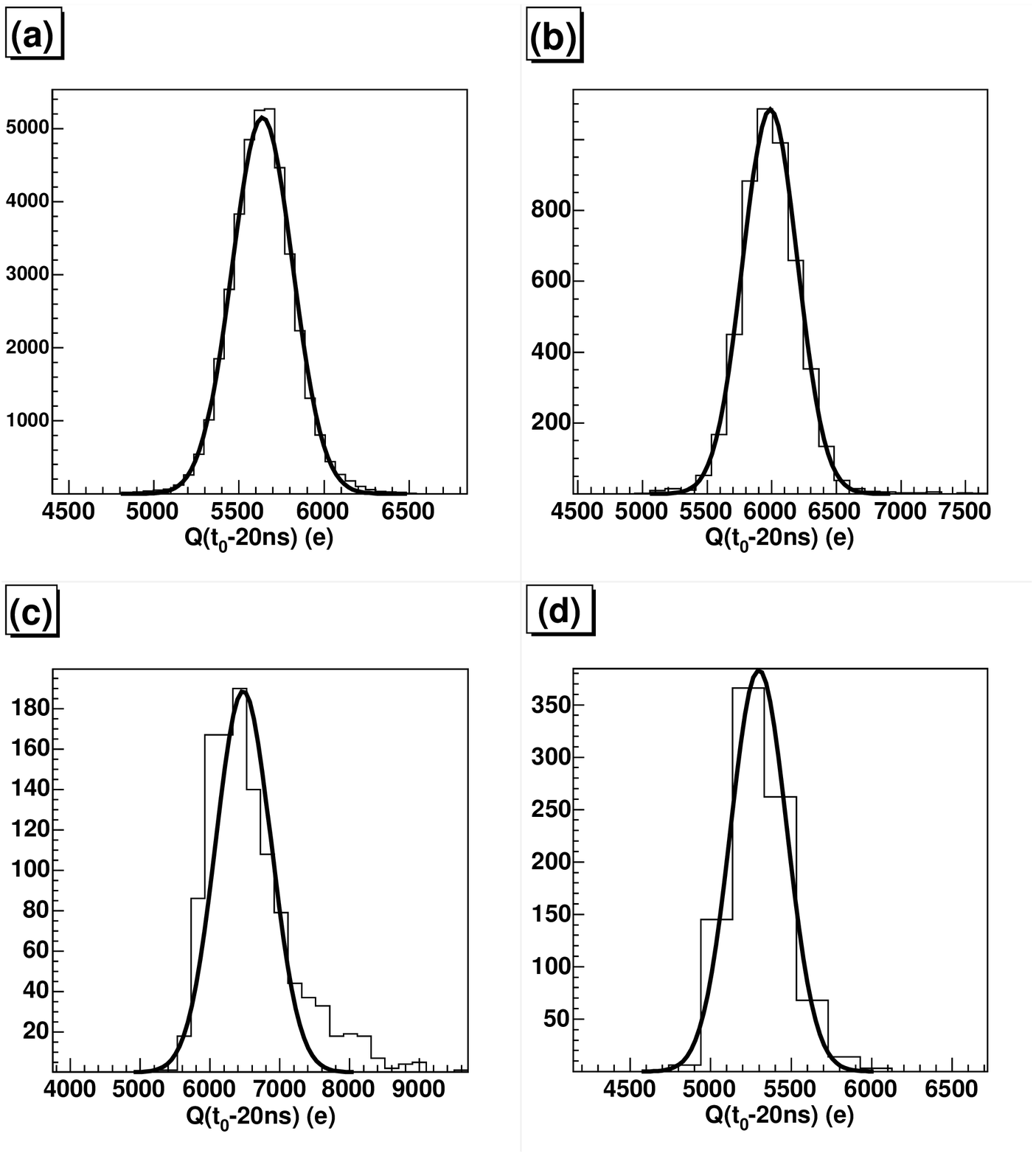}}
\caption{In-time-threshold distributions for the different pixel
types of a module tuned to an average threshold of 4,200~$e^{-}$,
(a) for standard pixel with a mean of 5,640~e$^{-}$ and
$\sigma=180~$e$^{-}$, (b) long pixel with a mean of 5,990~e$^{-}$
and $\sigma=210~$e$^{-}$, (c) ganged pixel with a mean of
6,680~e$^{-}$ and $\sigma=400~$e$^{-}$ and (d) inter-ganged pixel
with a mean of 5,300~e$^{-}$ and $\sigma=170~$e$^{-}$.
\label{fig:twalk}}
\end{figure}

A measurement of the timewalk, i.e. the variation in the time when
the discriminator input goes above threshold, is an issue since
hits with a low deposited charge have an arrival time later than
those with high charges, in particular for ganged pixels because
of their higher input capacity. The difference in threshold for a
signal arrival time of less than 20~ns and the nominal
discriminator threshold is for standard pixels
approx.~1,500~e$^{-}$, for ganged pixels approx.~2,300~e$^{-}$ and
for long pixels approx.~2,000~e$^{-}$, see figure~\ref{fig:twalk}.
Because the discriminator threshold can easily be tuned to values
below 3,000~e$^{-}$ the achieved timewalk is sufficient to meet
the ATLAS requirement of 6,000~e$^{-}$ for all pixels.

\clearpage

\begin{figure}[htb]
\begin{center}
\centerline{\includegraphics[width=\columnwidth]{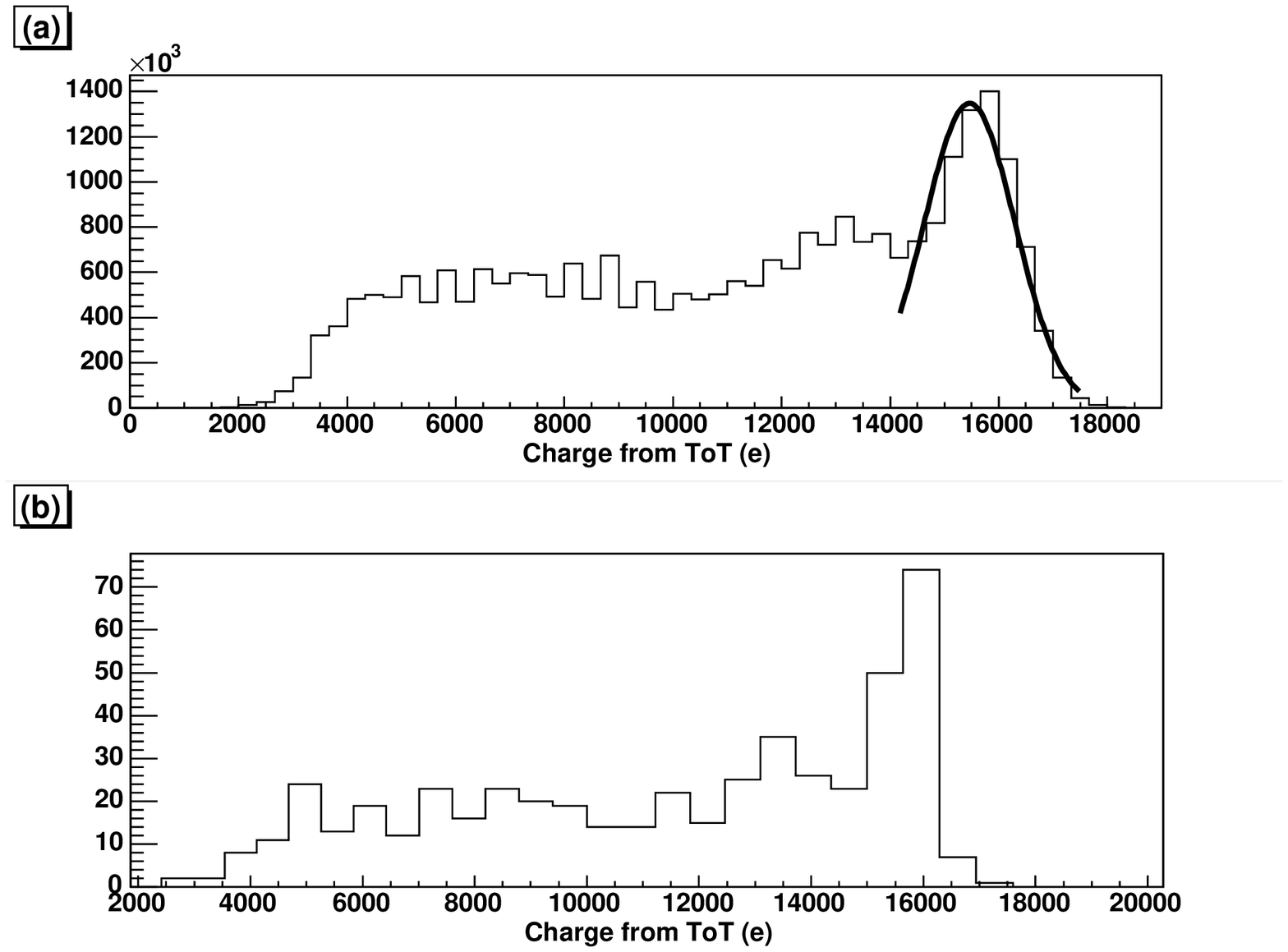}}
\caption{Am$^{241}$-Spectrum measured with an ATLAS pixel module
using the self-trigger capabilities and the ToT charge
information. Each channel of the module has been individually
calibrated. The spectrum (a) is a sum over all pixel without any
clustering. The peak has been fitted to 15,460~e$^{-}$. The
spectrum (b) is for a specific channel, i.e. chip 3, column 14,
row 33 given as an example. \label{fig:source}}
\end{center}
\end{figure}

Data taken when illuminating the sensor with a radioactive source
allows in-laboratory detection of defective channels. Such a
measurement obtained with an Am$^{241}$-source can be seen in
figure~\ref{fig:source}. $1,400,000$ events per FE-chips have been
accumulated for this measurement to ensure enough hits per channel
for a subsequent analysis. The integrated source-spectrum for all
pixels reconstructed from the ToT-readings is in agreement with
expectations (see figure~\ref{fig:source}, (a)); the main $60~$keV
$\gamma$ peak can clearly be distinguished from the background
which is dominated by events with charge sharing between
neighbouring pixels. Furthermore the individual pixel spectrum
(see figure~\ref{fig:source}, (b)) can be used for an absolute
charge calibration per readout channel, because here also the
$60~$keV $\gamma$ line can be identified.

\begin{figure}[htb]
\begin{center}
\centerline{\includegraphics[width=\columnwidth]{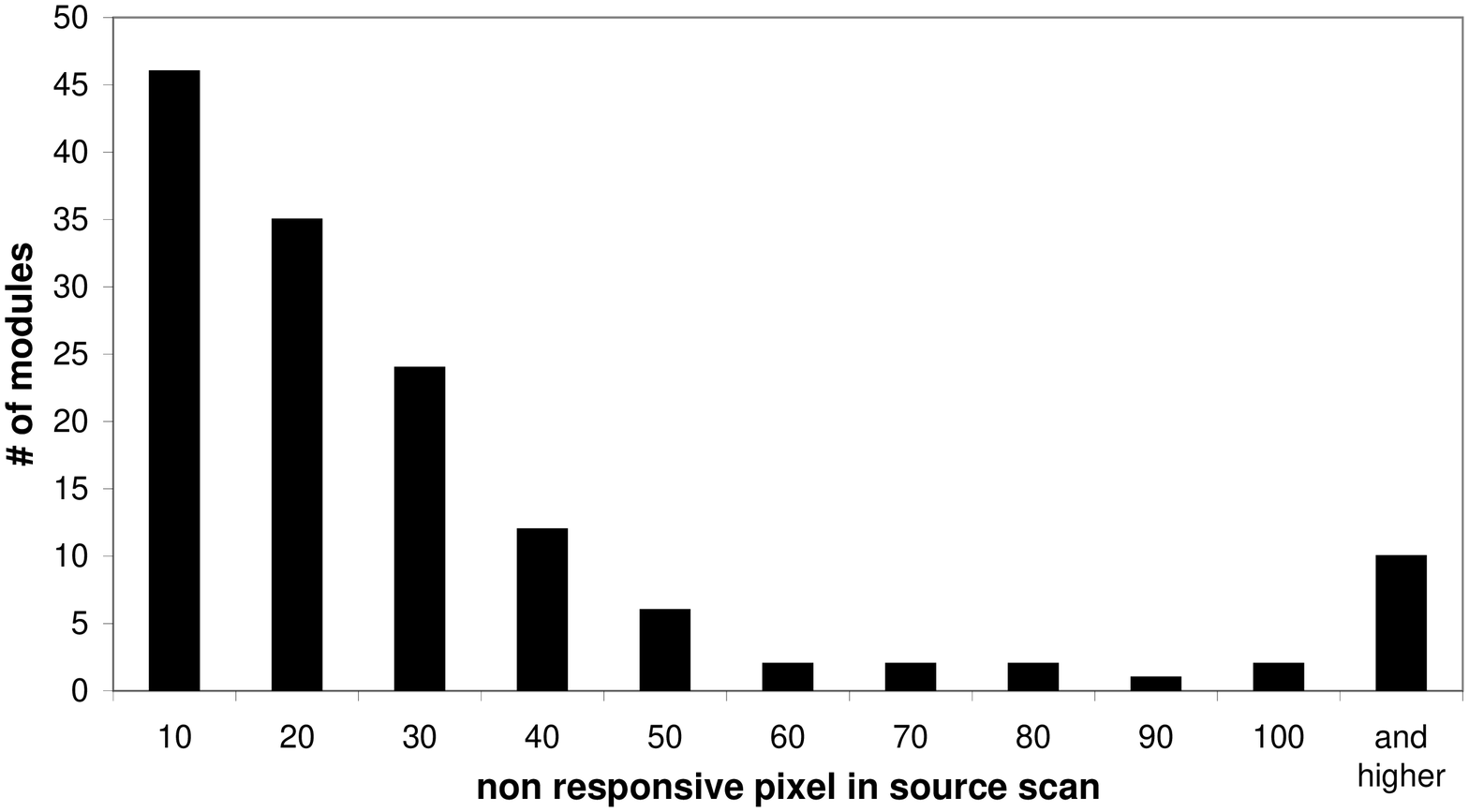}}
\caption{Distribution of the non responsive pixels for the first
150 modules produced for the ATLAS pixel
detector.\label{fig:yield}}
\end{center}
\end{figure}

Up to now roughly 150 modules have been produced and completely
characterized; every module undergoes an extensive test procedure
to ensure good performance inside the ATLAS detector. This
includes tests at room temperature as well as tests at the
operation temperature of $-10^{\circ}$C. A thermal cycling of at
least 48 hours with rapids cycles between $-30^{\circ}$C and
$+30^{\circ}$C to stress the modules is also part of the
procedure. Finally each module will be tuned and calibrated for a
source test to evaluate the number of non-responsive pixels. The
resulting distribution for the first 150 modules produced is shown
in figure~\ref{fig:yield}. Typically the number of defective
channels per modules is far less than 50 or $0.1$\% of all 46,080
pixels showing the excellent hybridization yield of the fine pitch
bump bonding.

\subsection{Test beam measurements}~\label{sec:testbeam}

\begin{figure}[htb]
\centerline{\includegraphics[width=.8\columnwidth]{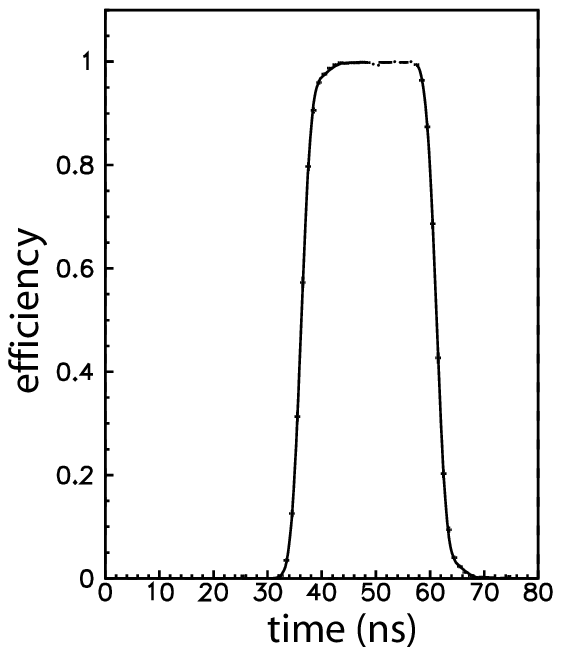}}
\caption{Efficiency vs. incident particle arrival time for an
ATLAS pixel module as measured in the test beam. At the flat top
an efficiency of 0.9990 is achieved. \label{fig:eff-unirr}}
\end{figure}

Tests have been performed in the beam line of the SPS at CERN
using 180~GeV/c hadrons. The setup consists of a beam telescope
for the position measurement~\cite{BAT}, trigger scintillators for
timing measurement to 36~ps, and up to four pixel modules. The
number of defective channels is observed to be less than
$10^{-3}$. For standard $50\times 400$~$\mu$m$^2$, non defective
pixels the efficiency for normal incidence particles is
99.90$\pm$0.15\% which can be seen in figure~\ref{fig:eff-unirr}.
Because the shown efficiency measurements contain also information
about the arrival time of the charge at the discriminator
w.r.t.~system clock the measurements allow also a determination of
the timewalk. The measured timewalk is in agreement with those
measurements from lab tests (see section~\ref{sec:lab}) giving a
timing window of 15~ns with high efficiency.

\begin{figure}[htb]
\centerline{\includegraphics[width=.8\columnwidth]{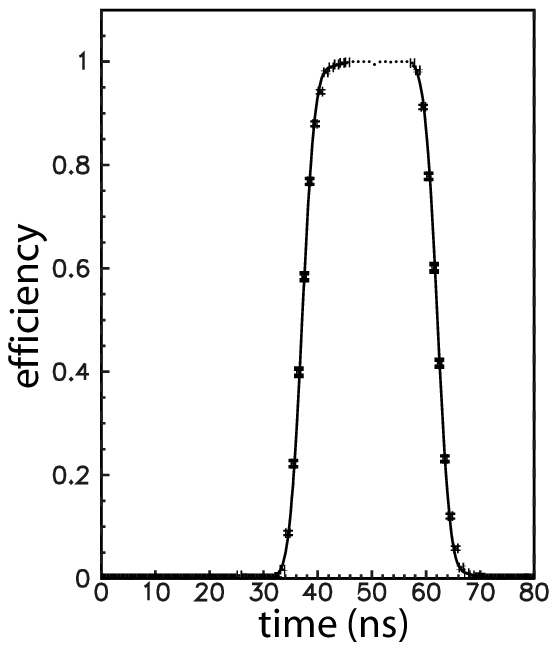}}
\caption{ Efficiency vs. incident particle arrival time for an
ATLAS pixel module in hit duplication mode as measured in the test
beam. At the flat top an efficiency of 1.0 is
achieved.\label{fig:eff-ideal}}
\end{figure}

Furthermore the efficiency of the ATLAS pixel modules can be
improved to perfect values of 100.00$-$0.03\% (see
figure~\ref{fig:eff-ideal}) by using a digital hit duplication of
the front end chip. Here the discriminator of each pixel
duplicates all hits below an adjustable ToT threshold to the
previous bunch crossing to recover the hit information for small
charges. Of course the drawback of this method is an increase of
the data volume inside the chip.

The space resolutions measured for one hit and two hit clusters
for different incident particle angles in binary readout mode,
i.e. approx. 12~$\mu$m in $r\phi$ and 110~$\mu$m in $z$ is
expected for the pixel size of $50\times 400~\mu$m$^2$. An
improvement of the resolution for two hit cluster can be achieved
by using a center of gravity method.

\subsection{Irradiation}~\label{sec:irr}

Seven production modules have been irradiated at the CERN PS with
24~GeV/c protons to a dose of 50~MRad ($1\cdot
10^{15}~$n$_{eq}$cm$^{-2}$), which is approximately the dose
expected after 10 years of ATLAS operation. The radiation damage
is monitored reading the leakage current individually for each
pixel. During irradiation the single event upset (SEU) probability
for the triple redundant pixel latches was measured by exposing
the full configured modules to the beam for several hours and then
read back the latches to search for bit flips. The achieved SEU
rate is of the order of $10^{-11}$ SEUs per proton for the $14$
pixel latches of each individual pixel cell showing no problems
with operation in such a harsch radiation environment.

\begin{figure}[htb]
\centerline{\includegraphics[width=.9\columnwidth]{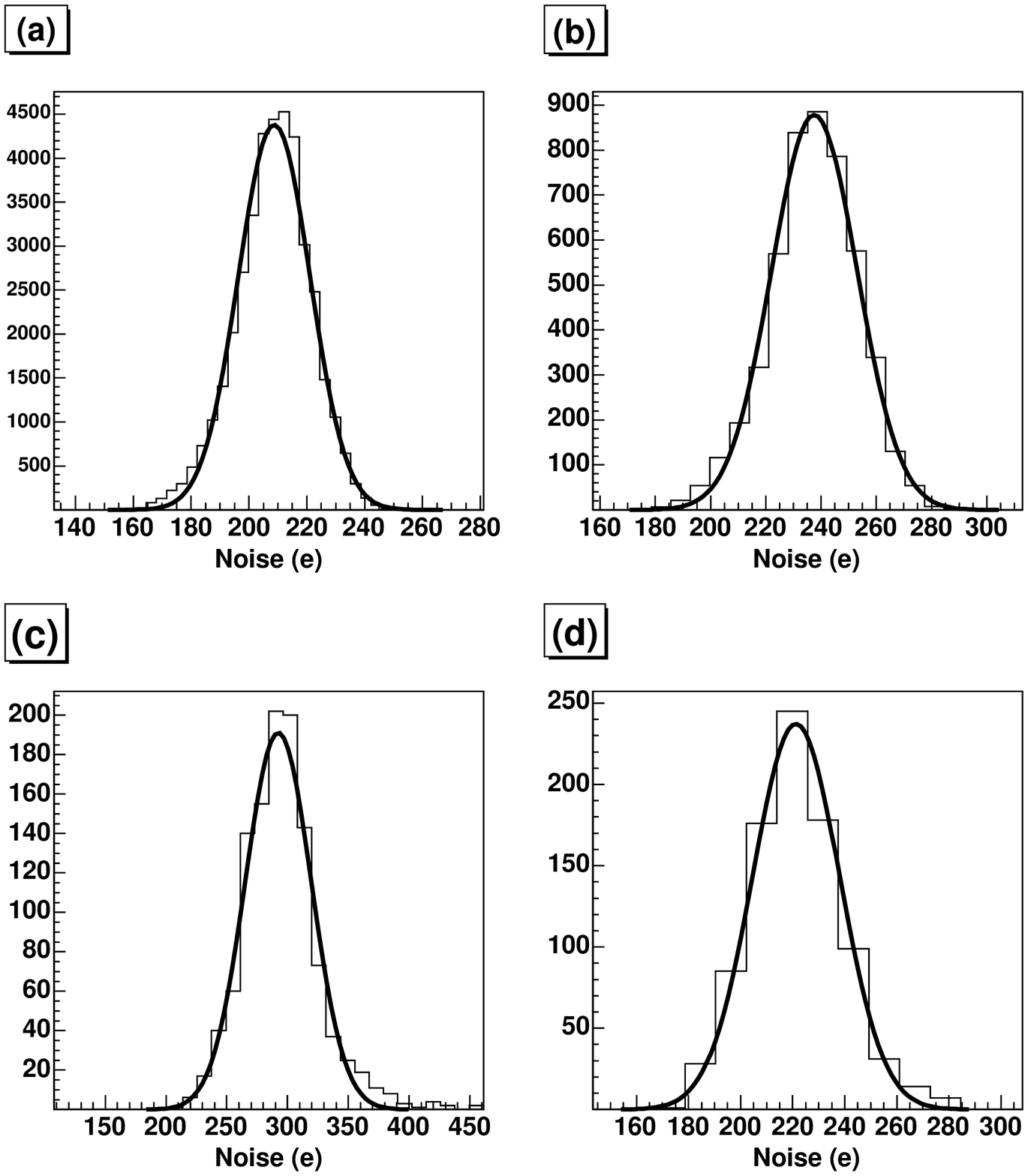}}
\caption{Noise distributions for the different pixel types of a
module irradiated with 24 GeV protons to a fluence of $1\cdot
10^{15}~$n$_{eq}$cm$^{-2}$, measured after re-tuning at
-4$^{\circ}$C; (a) for standard pixel with a mean noise of
209~e$^{-}$ and $\sigma=12~$e$^{-}$, (b) long pixel with a mean
noise of 238~e$^{-}$ and $\sigma=15~$e$^{-}$, (c) ganged pixel
with a mean noise of 292~e$^{-}$ and $\sigma=27~$e$^{-}$ and (d)
inter-ganged pixel with a mean noise of 221~e$^{-}$ and
$\sigma=17~$e$^{-}$. \label{fig:noise-irr}}
\end{figure}

The noise after irradiation as shown in figure~\ref{fig:noise-irr}
is only modestly increased and still well in agreement with
requirements for operation in ATLAS. Also the threshold dispersion
of such a highly irradiated module can be tuned to values of
60~e$^{-}$ as before irradiation.

\begin{figure}[htb]
\centerline{\includegraphics[width=.9\columnwidth]{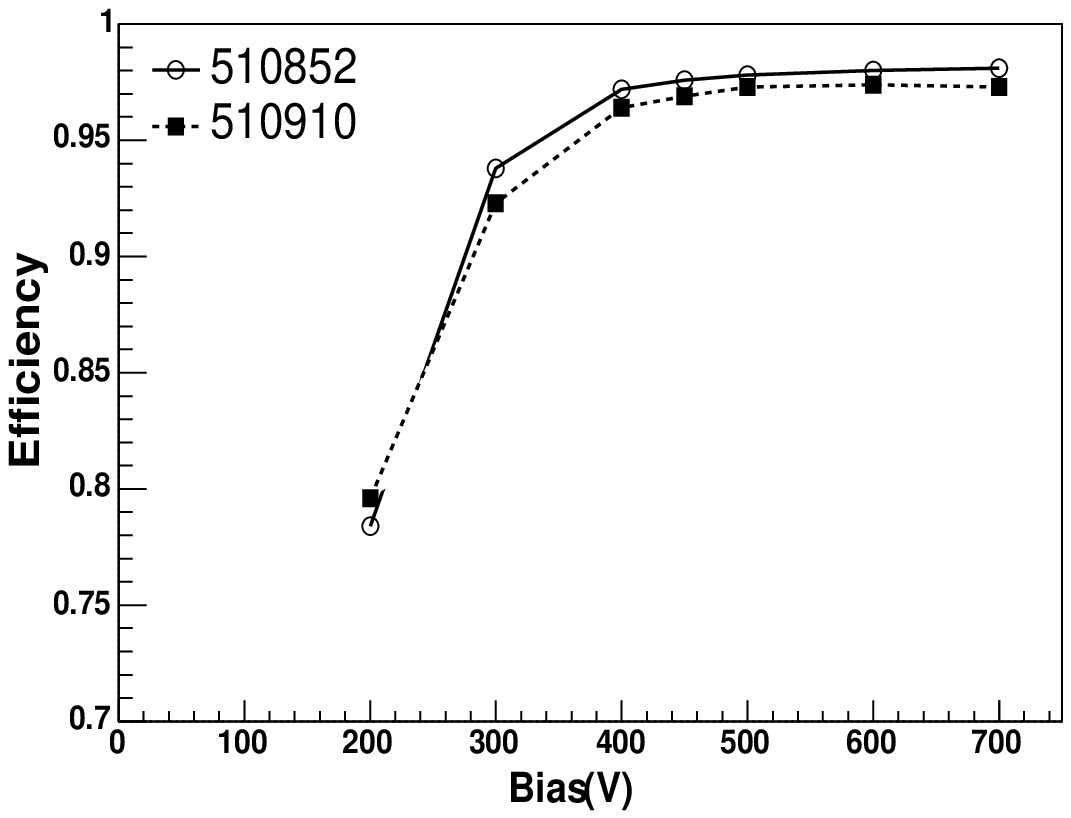}}
\caption{ Efficiency vs. bias voltage of two highly irradiated
pixel modules as measured in the beam
line.\label{fig:eff-vs-bias}}
\end{figure}

Irradiated modules have been tested in the beam line again (see
section~\ref{sec:testbeam}). The bias voltage needed for full
depletion has been measured for the highly irradiated modules
resulting to be between 400 and 500~V, see
figure~\ref{fig:eff-vs-bias}. This has to be compared with the
full depletion voltage of typically 50-80~V for modules before
irradiation. The deposited charge measured via the ToT readings
and the mean charge for irradiated modules is approximately
15,000~e$^{-}$ for a m.i.p. with an acceptable uniformity
w.r.t.~unirradiated modules.

\begin{figure}[htb]
\centerline{\includegraphics[width=.8\columnwidth]{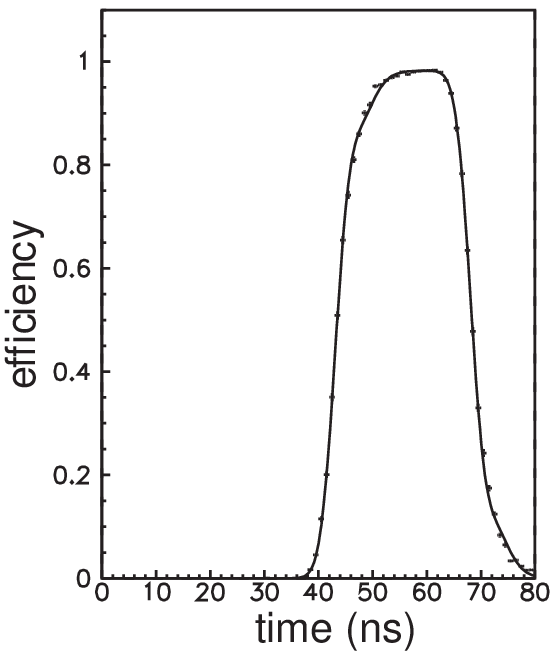}}
\caption{Efficiency vs. incident particle arrival time of an
irradiated module. At the flat top an efficiency of 0.9823 is
achieved.\label{fig:eff-irr}}
\end{figure}

Similar efficiency versus incident particle arrival time
measurements show for the highly irradiated modules efficiency
values of 98.23$\pm$0.15\%, well above the end-of-lifetime
requirement of 95\%, see figure~\ref{fig:eff-irr}. The slope of
the efficiency curve is slightly distorted w.r.t.~unirradiated
modules because of poor charge collection in a small region of the
irradiated sensor (``bias-dot'' region) which was implemented to
allow reasonable testing of the sensor without readout
electronics~\cite{sensor,mysensor}.

\section{Off-detector electronics}

The off-detector readout electronics is designed to process data
at a rate of up to 100~kHz level-1 triggers. The main
data-processing component is the ``read-out driver'' (ROD), of
which final prototypes have been built to pixel specifications and
are being evaluated. The first-step event-building and error
flagging is done via Field-Programmable-Gate-Arrays (FPGA). The
communication to the rest of the data acquisition system is run
through a 1.6~Gbit/s opto-link. The communication to modules,
online monitoring and calibration runs are performed with
Static-Random-Access-Memory(SRAM) and Digital-Signal-Processors
(DSP); their programming is ongoing and modules and small systems
have already been configured and operated successfully with a ROD.

All components of the off-detector electronics are in production
now and the progress and yields are well on track.

\section{System aspects}

\subsection{Support structures}

The mechanics of the system has to guarantee good positional
stability of the modules during operation while the amount of
material has to be kept to a minimum. At the same time it has to
provide cooling to remove the heat load from the modules and
maintain the sensors at a temperature of -$6^\circ$C to keep the
radiation damage low.

Barrel-modules are glued to ``staves'', long, flat
carbon-structures with embedded cooling pipes. The staves are
mounted inside halfshells, which themselves are assembled into
frames to form the barrel system.

The disks are assembled from carbon-sectors with embedded cooling
covering 1/8 of a wheel. The modules are glued directly to either
side of the disk sectors.

The module loading to staves and disk sectors requires high
position accuracy and good thermal contact without any risks to
damage modules during the process. First disk sectors and barrel
staves have been assembled with modules showing unchanged
performance of the individual modules after assembly.

The global support structures of the pixel detector are also made
of carbon structures and have been recently delivered. Currently
these structures are under test at CERN.
\subsection{System tests}

First system tests have been performed with six modules on a disk
sector and thirteen  modules on a barrel-stave. The noise
behaviour on the disks or staves shows no significant differences
compared to similar measurements with the same unmounted modules.
Larger system tests are already in preparation and will include
realistic powering and read-out.

\section{Conclusions}

Production modules built with the final generation of radiation
hard chips show largely satisfying performance in
laboratory-tests, in test beam studies and after irradiation.
Module production is well in progress with high yield and an
acceptable rate to finish the ATLAS pixel detector in time.

Work on the off-detector electronics and the support structures
have been going on in parallel and are well on track. First system
test results are promising.


\begin{thebibliography}{1}
\bibitem{IDTDR} Technical Design Report of the ATLAS Inner Detector,
CERN/LHCC/97-16 and CERN/LHCC/97-17 (1997).
\bibitem{PDTDR} Technical Design Report of the ATLAS Pixel Detector,
CERN/LHCC/98-13 (1998).
\bibitem{sensor} M.~S.~Alam et al. \emph{The ATLAS silicon pixel
sensors}, Nuclear Instr. Meth. A {\bf 456}, 217-232 (2001).
\bibitem{MCC} R.~Beccherle et al., \emph{MCC: the Module Controller Chip for the ATLAS Pixel Detector},
Nuclear Instr. Meth. A {\bf 492}, 117-133 (2002).
\bibitem{fabian} F.~H\"ugging, \emph{Front-End electronics and integration of ATLAS pixel modules},
accepted for publication in Nuclear Instr. Meth. A.
\bibitem{BAT} J.~Treis et al., \emph{A modular PC based silicon microstrip beam
telescope with high speed data acquisition}, Nuclear Instr. Meth.
A {\bf 490}, 112-123 (2002).
\bibitem{mysensor} F.~H\"ugging et al., \emph{Design Studies on
sensors for the ATLAS pixel detector}, Nuclear Instr. Meth. A {\bf
477}, 143-149 (2002).
\end{thebibliography}
\end{document}